\documentclass[amsmat,amssymb,amsfonts,aps,prb,twocolumn,showpacs]{revtex4}

\usepackage{graphicx}
\usepackage{dcolumn}
\usepackage{bm}
\begin{document}

\title{ Single Electron Transport in  electrically tunable nanomagnets}

\author{J. Fern\'andez-Rossier (1), Ram\'on Aguado (2)}
\affiliation{(1)
 Departamento de F\'{\i}sica Aplicada, Universidad de
Alicante, San Vicente del Raspeig, Spain
\\ (2) Instituto de Ciencia de Materiales de Madrid,CSIC,
Cantoblanco, 28049 Madrid,Spain}

\date{\today}

\begin{abstract}
We study a single electron transistor (SET) based upon a II-VI
semiconductor quantum dot doped with a single Mn ion. We present
evidence that this system behaves like a quantum nanomagnet whose
total spin and magnetic anisotropy depend dramatically both on the
number of carriers and their orbital nature.  Thereby, the
magnetic properties of the nanomagnet can be controlled
electrically. Conversely, the electrical properties of this SET
depend on the quantum state of the Mn spin, giving rise to
spin-dependent charging energies and hysteresis in the Coulomb
blockade oscillations of the linear conductance.


\end{abstract}

 \maketitle



Nanomagnets attract interest both because of their intriguing
behavior as relatively macroscopic quantum objects
\cite{Tejada} and their potential technological applications  as
magnetic bits \cite{Cobalt} and qbits \cite{Loss-Nature}.  The two
fundamental properties of a nanomagnet are the net spin of its
ground state, $S$ and its magnetic energy anisotropy tensor, ${\cal
K}$ that  governs the stability of the magnetization with respect
to quantum and thermal  fluctuations.  Although recent experiments
show that single molecule magnets like Mn$_{12}$ \cite{CM1,CM2} or
metallic Co \cite{Ralph-SET} nanoparticles can be probed in
single-molecule transistor measurements, their  properties can
hardly be tuned once they are fabricated. Here we show that  a
single electron transistor (SET) consisting of a II-VI quantum dot
doped with a single Mn atom behaves like a tunable nanomagnet
whose magnetization and anisotropy axis can be  reversibly
manipulated electrically. Conversely, the  conductance and charging energy of
the tunable nanomagnet depend on the quantum state of
the Mn spin and are  not uniquely
determined by the gate and the bias voltage.

Our proposal is based on  two independent progress in nanofabrication. On one
side, the fabrication and optical probing of single CdTe quantum dots doped
with {\em a single Mn atom}
\cite{Besombes1,Besombes2,Besombes3}. In the absence of carriers,  the spin
$S=5/2$ of the Mn is free.  Optical excitation of electron-hole pairs into the
dot shows that the Mn spin is exchange coupled
to both the electron and the hole\cite{Besombes1,Besombes2,Besombes3,JFR05}.
%
On the
other side, the control of the charge state of II-VI semiconductor quantum dots
with single electron accuracy has been experimentally demonstrated
\cite{Klein-Nature,Forchel-APL} as well as in the case of single-Mn doped
quantum dots \cite{Leger-2006} and Mn-doped GaAs islands \cite{Wunderlich}.

{\em Hamiltonian}.
We consider a CdTe quantum dot (QD) doped with a single  Mn, weakly coupled to
two metallic and non-magnetic electrodes.
 The dot can be gated so that either the valence band or the
conduction band is in resonance with the metallic reservoir and the number of
either electrons or holes is varied at will.
The total Hamiltonian reads:
${\cal H}={\cal H}_{QD}+{\cal H}_C+{\cal H}_L + {\cal H}_R + {\cal
V}_L+{\cal V}_R.$
Here ${\cal H}_{QD}$ is the Hamiltonian for the diluted magnetic
semiconductor (DMS) quantum dot. In analogy with the the standard
model \cite{Furdyna} for bulk DMS, ${\cal H}_{QD}$ describes
confined conduction band electrons and valence holes
interacting with a localized Mn
spin $S=\frac{5}{2}$, denoted as $\vec{M}$, via a local exchange interaction.
 QD carriers occupy
localized  spin orbitals $\phi_{\alpha}$ with energy $\epsilon_{\alpha}$ which
are described in the envelope function $\vec{k}\cdot\vec{p}$ approach
 \cite{JFR04,JFR05,dot-holes}. In the case of valence band holes
the  6 band Kohn-Luttinger Hamiltonian, including spin orbit interaction, is
used as a starting point to build the quantum dot states \cite{dot-holes}.
The second quantization Hamiltonian of the isolated dot describes the states of reads:
\begin{eqnarray}
{\cal H}_{QD}=
\sum_{\alpha,\alpha'}
\left(
\epsilon_{\alpha}
 \delta_{\alpha,\alpha'}
 +
J_{\alpha,\alpha'}
\vec{M}\cdot
\vec{S}_{\alpha,\alpha'}
\right)f^{\dagger}_{\alpha}f_{\alpha'}
\label{hamil}
\end{eqnarray}
Here $f^{\dagger}_{\alpha}$ creates  a band carrier in the
$\alpha$ single particle state of the quantum dot, which can be
either a valence band or a conduction band state. The first term
in the Hamiltonian describes non interacting carriers in the dot
and the second term describes the exchange coupling of the
carriers and the Mn. We neglect  interband exchange so that
$J_{\alpha,\alpha'}=J_e$ ($J_{\alpha,\alpha'}=J_h$) if both
$\alpha$ and $\alpha'$ belong to the conduction band states
(valence band states). In contrast, we include
exchange processes by which a
carrier is scattered between two different levels of the dot that
belong to the same band.
The matrix elements of both valence and conduction band spin density,
evaluated at the location of the Mn atom, are given by
$\vec{S}_{\alpha,\alpha'}$. They
depend strongly on the orbital nature of the single particle level in question.
In the case of conduction band we neglect spin orbit interactions so
that $\vec{S}_{\alpha,\alpha'}$ is rotationally invariant
\cite{JFR04}. In contrast,  strong spin orbit interaction of the
valence band makes the Mn-hole interaction strongly anisotropic
\cite{JFR05,Leger-2006,dot-holes} and it varies between different dot levels.
Following previous work \cite{JFR05,JFR04,dot-holes}
confinement is described by a hard wall cubic potential
 with   $L_z<L_x\simeq L_y$. Although  real dots are not cubic,
 this simple model \cite{dot-holes,JFR05}
 provides an excellent description of
the Hamiltonian of the Mn spin coupled to the carriers, which is able
to account for the
 non-trivial single-exciton PL spectra both for neutral
\cite{Besombes1,Besombes2,Besombes3} and charged\cite{Leger-2006}
 single-Mn doped CdTe QD.

Coulomb repulsion between carriers is described within the
constant interaction model \cite{Beenakker-REV}:
${\cal H}_C=\frac{1}{2C}({\hat Q}+C_gV_g+C_L\frac{\mu_L}{e}+C_R\frac{\mu_R}{e})^2$,
where $C=C_L+C_R+C_g$ is the total capacitance to the external
circuit, $C_L$ and $C_R$ are the capacitances of the left and
right junctions ($eV_B\equiv \mu_L-\mu_R$ is the bias voltage) and
$C_g$ is the capacitance to the gate (with voltage $V_g$). ${\hat Q}$ is
the extra charge in the dot. We do not consider dots with orbital
degeneracy for which Coulomb correlations, neglected in this
paper,  are relevant \cite{Govorov05a,PRL2005-Haw}. Finally,
${\cal H}_L=\sum_{\sigma,k} \epsilon_k
a^{\dagger}_{k\sigma}a_{k\sigma}$ and ${\cal H}_R=\sum_{\sigma,p}
\epsilon_p b^{\dagger}_{p\sigma}b_{p\sigma}$ describe the metallic
electrodes and ${\cal V}_L= \sum_{\sigma,k,\alpha}
V_{\sigma,k,\alpha}f^{\dagger}_{\alpha}a_{k\sigma} +h.c$ and
${\cal V}_R= \sum_{\sigma,p,\alpha}
V_{\sigma,k,\alpha}f^{\dagger}_{\alpha}b_{p\sigma} +h.c$ are the
standard spin-conserving tunneling Hamiltonian that couple the
metallic reservoirs and the dot.

We first discuss the properties of the {\em eigenstates}
$|N\rangle$ of ${\cal H}_{QD}$ for isolated dots (${\cal
V}_L={\cal V}_R=0$) with a given number of carriers, interacting
with the Mn atom. We show results for two dots of CdTe with $L_z=
60\AA$, $L_x=80 \AA$ and different $L_y=80\AA$ (dot A) and
$L_y=75\AA$ (dot B), both doped with 1  Mn atom. The neutral dot
has 6  degenerate states, corresponding to  the $(2S+1)$
equivalent spin orientations  of the $S=5/2$ Mn spin.  This
degeneracy is lifted in the presence of either electrons or holes.
We focus on dots with a odd number of carriers (open shells) for
which the interactions are stronger \cite{JFR04,Govorov05a} and
study how the magnetic {\em anisotropy} varies with the number of
carriers. The spectra of dots with 1 electron, 1 hole and 3 holes
are shown in figs. 1a, 1b and 1c respectively for dot A (and also
QD $B$ for the case of 1 hole).  The effect of intra-level exchange
is  magnified in the inset of fig. 1a. In figure 1b and 1c we only
show the low energy manifold for  dot A with 1 and 3 holes. The 12
states of the low energy  manifold for
$Q=\pm 1$
 and $Q=+3$ are formed mainly by the two lowest energy electronic
configurations of the dot with a single unpaired fermion coupled
to the 6 Mn states. The low energy sector of ${\cal H}_{QD}$ can
be described by an intra-level effective Hamiltonian\cite{Leger-2006}:
${\cal H}_{eff}=
j^x \tau^x M^x+
j^y \tau^y M^y+j^z \tau^z M^z$
where $\tau^a$ are the Pauli matrices operating on the isospin space
defined by the lowest energy single particle
doublet.

Both the absolute and the
relative values of $j_x$, $j_y$ and $j_z$ depend  mostly on the spin
properties of the external shell of the quantum dot: either conduction band
level ($ Q=-1$), heavy hole ($Q=+1$) or light hole ($Q=+3$).
 Thereby, the $j_{x,y,z}$ can be controlled reversibly by means of the
gate voltage in the same  device. The effective Hamiltonian of the
Mn coupled to the "master fermion" in dot A goes  from ferromagnetic Heisenberg
($j_x=j_y=j_z<0$) when ($Q=-1$) to antiferromagnetic Ising $j_x=j_y=0,
j_z>0$ when ($Q=+1$) to XXZ ($j_x=j_y>j_z$). In dot $B$ similar results
are obtained, with $j_x\neq j_y$, which provides a spin-flip term in the
$Q=+1$ case, absent in dot A.
\begin{figure}
[t]
\includegraphics[width=3.0in]{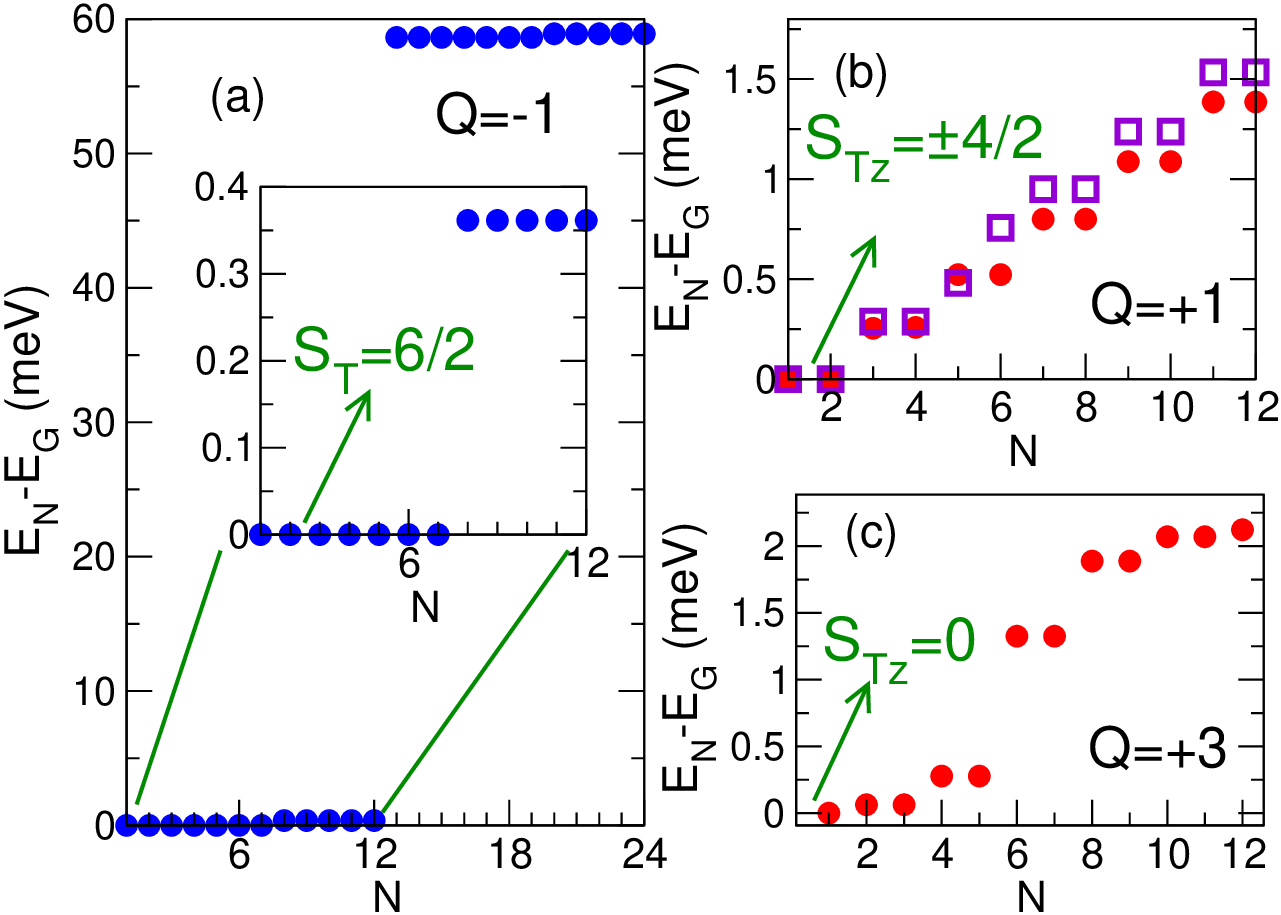}
\caption{ \label{fig1}(Color online). Low energy spectra of ${\cal
H}_{QD}$ for dot A (circles) with ${\cal Q}=-1$ (a),$Q=+1$
(b) and $Q=+3$ (c). Notice the different vertical scale in
(a) and (b) In panel (b) we also show the spectrum of QD $B$ with
$Q=+1$ (squares). The Kohn-Luttinger parameter for CdTe are
$\gamma_1=4.14$, $\gamma_2=1.09$ and $\gamma_3=1.62$ and the spin
orbit interaction is $\Delta=950$ meV.  We take $J_e= -15$ eV
$\AA^3$ and $J_h= 60 $eV $\AA^3$. }
\end{figure}

Correspondingly, the spin properties  of the ground state also change as  a
function $Q$. In the case of $Q=-1$ the Mn spin $5/2$ and the conduction
band electron couple ferromagnetically to yield a septuplet with $S_T=3$.
In the case of $Q=+1$ the ground state doublet corresponds to the Mn
spin maximally polarized against the heavy hole spin,
$|M_z=-5/2,\uparrow \rangle$ and $|M_z=+5/2,\downarrow \rangle$, both for dots
A and B. In dot A the rest of the low-energy sector is formed by 5 doublets
eigenstates of both $M_z$ and $\tau^z$, whereas in dot $B$ the small spin-flip
interaction mixes
$|M_z=+1/2,\downarrow \rangle$ and $|M_z=-1/2,\uparrow \rangle$.
In the case with $Q=+3$, the ground state is not degenerate
and the Mn spin is polarized in the $xy$ plane, minimizing $M_z$.
These
differences reflect the spin properties of conduction band electron, heavy hole
and light hole respectively.

We now address how these remarkably different magnetic properties
occurring in the same  dot are reflected in the electrical
behavior of the SET. In analogy with previous work
\cite{Efros,Waintal-PRL2003,Waintal-PRL2005,Timm,Braun1},
 we derive a  quantum master
equation for the dissipative dynamics of the {\em reduced density
matrix} $\rho_{NM}(t)$ written in the basis of many-body states
$|N\rangle$.
  Importantly, this quantum master equation includes
the combined dynamics of both populations and coherences. The
latter are important because of the intrinsic many-body
degeneracies of the QD spectra shown in fig. 1. Assuming that the
quantum dot is weakly coupled to the electronic reservoirs
(sequential tunneling), the dissipative dynamics of
the density matrix is governed by a Markovian kernel,
$\dot{\rho}(t)=A\rho(t)$, where $\rho$ can be casted as
a  vector containing
both  populations and coherence terms.
The matrix $A$ contains information about dissipative dynamics of $\rho$ 
which is governed by the rates:
%
\begin{eqnarray}
\Gamma_{N,M}^{\pm} &=&\sum_{r\in L,R}\Gamma_r
n_r^{\pm}(E_N-E_M)\sum_\alpha |\langle N|f^{\pm}_{\alpha} |M
\rangle|^2 \label{gamma}
\end{eqnarray}
Here $f^{+}_{\alpha}\equiv f^{\dagger}_{\alpha}$,
$f^{-}_{\alpha}\equiv f_{\alpha}$, $n_r^+$ is the Fermi function
of reservoir r and $n_r^-$=1-$n_r^+$.The notation
$\Gamma_{N,M}^\pm$ implies that states $M$ with charge $Q$
are connected with states $N$ with charge $Q\pm 1$. . The
coupling to the leads is parametrized by
$\Gamma_{L,R}=\frac{2\pi}{\hbar}|V_{L,R}|^2 N_{L,R}$, where
$N_{L,R}$ is the DOS of the metallic reservoir.
Once we obtain the {\em steady state} density matrix $\tilde{\rho}$ (namely,
$A\tilde{\rho}=0$),  we can compute the average charge,
magnetization and current.
To lowest order in $\Gamma_{L,R}$, the most general expression for
the current can be written as $I=\frac{I_L-I_R}{2}$ with:
\begin{widetext}
 \begin{eqnarray}
I_{L/R}=e\Gamma_{L/R}\sum_{N,N'}\sum_{M}\sum_\alpha
\tilde{\rho}_{N,N'}\{n_{L/R}^+(E_N-E_{N'})  \langle N'|f_{\alpha}
|M \rangle\langle M|f^{\dagger}_{\alpha} |N \rangle
-n_{L/R}^-(E_N-E_{M})  \langle N'|f^{\dagger}_{\alpha} |M
\rangle\langle M|f_{\alpha} |N \rangle\}\label{current}
\end{eqnarray}

Notice that Eq. (\ref{current}) includes both diagonal and
non-diagonal terms in the density matrix. The latter are important
when two degenerate states with $Q$ are coupled to the same
state of $Q\pm 1$ via a single tunneling event.
\end{widetext}

The steady state of a standard SET is uniquely characterized by
external voltages \cite{Beenakker-REV}. For instance, a new charge
is accommodated in the dot at precise values of the gate voltage,
when the electrochemical potential of the dot $\mu(N)$ (the energy
required for adding the $N$th electron to the dot) falls within
the bias window $\mu_L\geq\mu(N)\geq\mu_R$. When this condition is
met, the number of electrons can vary between $N-1$ and $N$
resulting in a single-electron tunneling current.
Importantly, in
our case the charge and the conductance of the SET \emph{depend
also on the quantum state of the Mn spin}.

In figure 2 we show  linear conductance $G_0(V_G)$, average charge and diagonal
terms of the steady state $\rho$, as the gate produces the transition between
charge zero and chage $\pm 1$ for both  electrons (left panels) and holes
(right panels).  The initial $V_G$ for the charging simulations is chosen so
that only the  ${\cal Q}=0$ states are occupied. This initial condition is
described by a thermal  $\rho$  with 6 equally populated Mn spin states,
$M_z=\pm 5/2,\pm 3/2,\pm 1/2$. We ramp the gate and solve the master equation 
to obtain the {\em steady state} $\rho$, which is used as initial condition for
the next run with higher $V_G$. In the case of electrons we obtain standard
results: a single peak in the $G_0(V_G)$ curve occurs as the gate is ramped
to change the charge of the dot by one unit.
In fig. 2c we show the evolution of the steady state populations:  The 6 ${\cal
Q}=0$ spin states are relaxed altogether in favor of the 7 states of the ${\cal
Q}=1$ and $S_T=3$ states.

The results for holes in QD $A$ are remarkably different:
in the process of injection of 1
hole the $G_0(V_G)$ curve shows 3 peaks instead of 1.
This results from the lack of Mn spin relaxation ($M_z$ is a
conserved quantity for the entire Hamiltonian including tunneling),  
which makes the steady  $\rho$ different from  to the thermal $\rho$.
As the gate brings down into
resonance the 2 ${\cal Q}=+1$  ground states ($+5/2,\downarrow)$ and
($-5/2,\uparrow$)
with 
the 6 ground states with ${\cal Q}=0$,
population tranfer only affects states with $|M_z|=5/2$ in
both charge sectors (first peak in fig. 2d at $V_G=4.3$meV). 
Further increase of
the gate brings the energy of the 
the  ${\cal Q}=+1$ ground state doublet below the 
${\cal Q}=0$ states with  $|M_z|\neq 5/2$, which are not depleted because
 $M_z$   is conserved (fig. 2f).  The population transfer only occurs
 when the second and
third doublet
of the ${\cal Q}=+1$ spectrum, with $|M_z|=3/2$ and $|M_z|=1/2$ are brought
down in resonance with the ${\cal
Q}=0$ states. This accounts for the other two peaks in the $G_0(V_G)$ curve as
the charge of the dot approaches $+1$. Hence, the   charging energy for holes
depends on the absolute value of the  {\em spin} of the Mn. 
The discharge simulation is done  analogously.  If the initial
$V_G$ is such that there is one hole in the QD, the Ising
interaction removes the degeneracy among states with different
$|M_z|$. Only the doubly degenerate ground state of the ${\cal Q}=+1$ 
sector is occupied in thermal equilibrium. As the
gate is ramped to discharge the dot, a single peak in the
conductance is obtained, corresponding to the resonance condition
with between the ${\cal Q}=0$ and ${\cal Q}=+1$ states with $|M_z|=5/2$ .

\begin{figure}
[b]
\includegraphics[width=3.15in,height=2.1in]{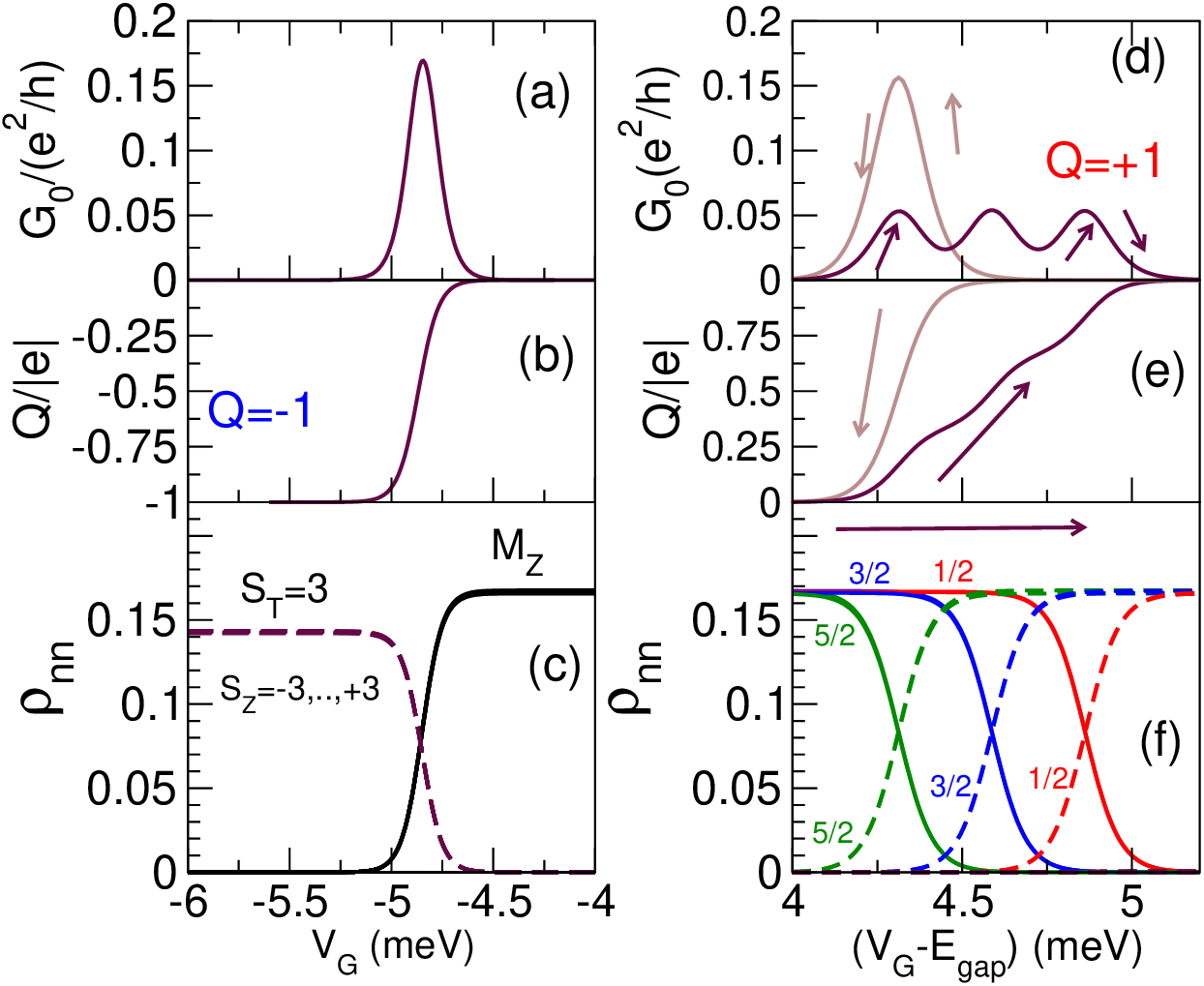}
\caption{ \label{fig2}(Color online). $G_0(V_G)$ (upper panels),
 charge (middle panels) and diagonal terms of
the $\rho$ (lower panels)   for
QD $A$ as a function of $V_G$ around the $Q=-1\leftrightarrow0$
transition (left) and  $Q=0\leftrightarrow+1$  (right).
Lower panels: solid (dashed) lines correspond to ${\cal Q}=0$ 
($|{\cal Q}|=1$) states. 
Results obtained with
$\Gamma_L=\Gamma_R=0.01$meV and $k_BT=$0.05 meV.}
\end{figure}

The difference between electrons and holes arises from 
the different value of a crucial time scale in the magnetic sigle electron
transistor:
the Mn spin relaxation time, $T_1$. In the case of QD $A$,
$T_1$ is infinite for holes (Ising coupling) 
which makes the steady state  different from the
thermal state. In the case of electrons (left panels)  the transverse spin
interactions make $T_1$  comparable to
the charge relaxation time ($\Gamma_{L,R}^{-1}$) so that steady and thermal 
$\rho$ are identical. In real dots doped with one hole, $T_1$ may be long but not
infinite. Two independent mechanisms, missing in the simulations shown in the
right panels of fig. 2.,  yield a finite $T_1$.  First, the Mn $T_1$
due to super-exchange with other $Mn$ spins which scales exponentially with the Mn
density \cite{T1Mn}. For bulk Cd$_{0.995}$Mn$_{0.005}$Te  we have
$T_1=10^{-3}s$, which is a lower limit estimate  for $T_1$ of  the QD with a
single Mn. The second mechanism is the small\cite{Leger-2006}
transverse spin interaction, which is
proportional to the light-hole heavy-hole mixing. 
We have simulated QD $B$, for  which
spin-flip interaction between the hole and the Mn is small but non-zero
 resulting in a finite 
$T_1$. If we integrate the master equation for
$\Gamma^{-1}<<t<<T_1$ the $G_0(V_G)$ curve displays 2 peaks and hysteretic
behaviour. In contrast, if we integrate the master
equation for $\Gamma^{-1}<<T_1<<t$, the system reaches the
equilibrium state for each value of $V_G$  so that the $G_0(V_G)$ 
  curve has a single
peak. Therefore, we claim that
effects related to incomplete spin
relaxation of the dot will be observed subject to two conditions: 
the  finite bandwith of the measurements should be larger than $1/T_1$ (see for
instance ref. (\onlinecite{CM2}))
and the pace at which $V_G$ is ramped should be faster than $T_1$.

\begin{figure}
[t]
\centering %
\includegraphics[width=3.0in,height=2.0in]{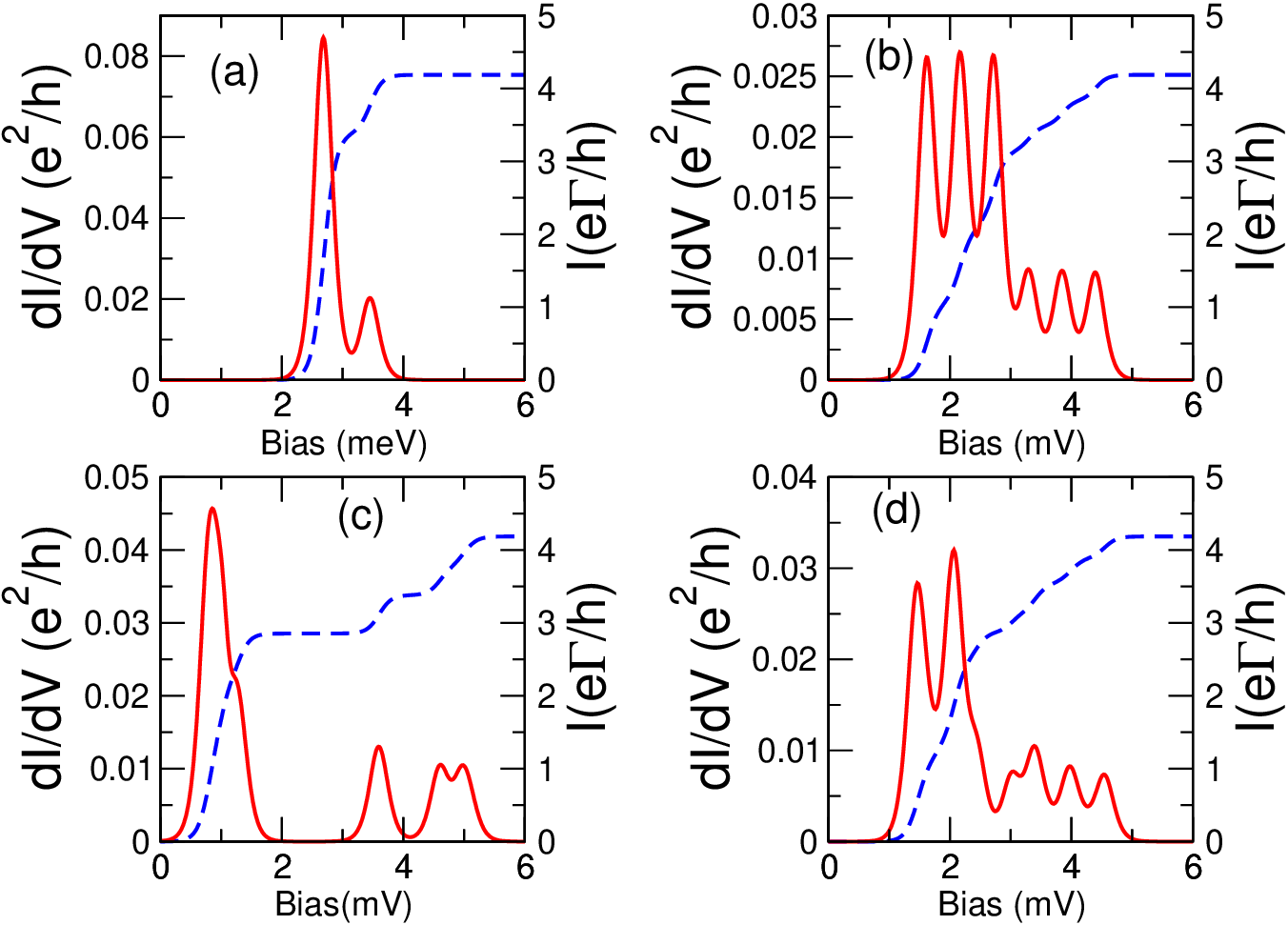}
\caption{ \label{fig3}(Color online). Current and differential
conductance as a function of bias for QD $A$ (a,b,c) at various
charge states and QD $B$ (d) (see text).}
\end{figure}

The finite bias conductance of the device also
depends strongly on the charge state  of the dot.
In figure 3 we show $I(V_B)$ and $\frac{dI}{dV_B}$ curves for  dot
$A$ corresponding to bias-assisted single electron fluctuations
between $Q=0$ and $Q=-1$  (fig 3a), $Q=0$ and
$Q=+1$   (fig. 3b), and $Q=2$ and $Q=+3$ (fig
3c). Fig. 3d is the analogous of 3b for dot B. Current flows
whenever the addition of a fermion is permitted by energy
conservation and spin selection rules. The former provides a link
between the $\frac{dI}{dV_B}$ curve and the energy spectra of dots
shown in fig. 1 (since the spectra of dots with $Q=0$ and
$Q=2$ are flat). 
Interestingly, the $\frac{dI}{dV_B}$ for electron
tunneling (3-a) shows a
zero-magnetic field splitting related to recent experimental observations
\cite{Gould06}.
In turn,   $\frac{dI}{dV_B}$ has
6(7) peaks in fig. 3b (3d), are  similar to the  experimental
single exciton PL spectra \cite{Besombes1,Besombes2,Besombes3,JFR05}.

In summary, we have shown some of the equilibrium and non-equilibrium
properties of a semiconductor quantum dot doped with a single Mn atom and wired
as a single electron transistor. The different orbital nature of the conduction
band electrons, heavy holes and light holes  determines both the total spin and
the magnetic anisotropy of the dot. In the case of holes, for which Mn spin
flip processes are  heavily inhibited, we predict different results for the
$G_0(V_G)$ curves depending on whether the system is relaxed to equilibrium or
not. In the case of the latter, we predict  hysteretic Coulomb blockade
oscillations related to the  the quantum state of the Mn spin. Because most of
the transport properties discussed above are inherent to nanomagnets with long
spin relaxation time, our findings  might be very general and have implications
in recent experiments \cite{CM1,CM2,Wunderlich}.

Fruitful discussions with L.  Brey, J. J. Palacios, Y. L\'eger, L.
Besombes, J.  Cibert,  H. Mariette, C. Gould and P. Hawrylak are acknowledged.
 This work
has been financially supported by MEC-Spain (Grants FIS200402356,
MAT2005-07369-C03-03, and the Ramon y Cajal Program) and by
CAV (GV05-152).


\widetext
\end{document}